\documentclass[a4paper,12pt]{extarticle}
\usepackage[utf8]{inputenc}
\usepackage{cmap}
\usepackage[english, russian]{babel}
\usepackage{amsthm,amssymb,amsmath,amsfonts,latexsym,enumerate,float}
\usepackage[left=20mm,right=20mm,top=2cm,bottom=2cm]{geometry}
\usepackage[dvips,pdftex]{graphicx}
\graphicspath{{pictures/}}
\usepackage{color}
\usepackage{cite}
\usepackage{tempora} 
\usepackage{indentfirst} 
\usepackage{tocloft} 
\usepackage[left]{lineno} 
\usepackage{authblk}

\usepackage{titlesec}
\titleformat{\section}[block]{\Large\bfseries\centering}{\thesection}{1ex}{}

\usepackage{setspace} 
\usepackage{wrapfig}
\usepackage{enumitem}
\addto\captionsrussian{}

\title{Simple Realistic Model of Spin Reorientation in 4f-3d Compounds}
\date{}
\author{Alexander Moskvin\thanks{alexander.moskvin@urfu.ru}, Evgenii Vasinovich, Anton Shadrin}
\affil{Ural Federal University, Ekaterinburg, Russia}

\begin{document}
\def\figurename{Fig.}
\maketitle

\textbf{Abstract:} Spin reorientation is an important phenomenon of rare-earth perovskites, orthoferrites and orthochromites. In this study, we consider a simple but realistic microscopic theory of the spontaneous spin-reorientation transitions induced by the 4f-3d interaction,  more specifically, the interaction of the main Kramers doublet or non-Kramers quasi-doublet of the 4f ion with an effective magnetic field induced by the 3d sublattice. The obtained results indicate that the cause of both the temperature and the character of the spin-reorientation transition is a competition between the second and fourth order spin anisotropy of the 3d sublattice, the crystal field for 4f ions, and the 4f-3d interaction.

\textbf{Keywords:} 4f-3d interaction; (quasi)doublets; spin reorientation

\section{Introduction}

Rare-earth orthorhombic perovskites, orthoferrites \textit{R}FeO$_3$ and orthochromites \textit{R}CrO$_3$ (where \textit{R} is a rare-earth ion and yttrium), exhibit many important features such as weak ferro- and antiferromagnetism, magnetization reversal, anomalous circular magnetooptics, and the phenomenon of the spontaneous spin reorientation. The spin reorientation (SR) is one of their unique properties that have attracted a lot of attention back in the 70s of the last century\,\cite{BZKL,KP}, though their exact microscopic origin is still a challenge to theorists and experimentalists.

The revival of interest in the mechanism of the spontaneous spin reorientation and magnetic compensation in rare-earth perovskites in recent years is related with the discovery of the magnetoelectric and the exchange bias effect, which can have a direct application in magnetoelectronics. Along with the emergence of new experimental studies (see, e.g., Refs.\,\cite{PRB102,PRB103}), there also appeared theoretical works claiming to modify the mean-field theory of the spontaneous spin-reorientation transitions\,\cite{Bazaliy} or to scrutinize the microscopic mechanism responsible for spin reorientations and magnetization reversal\,\cite{Luxem}. In fact, these results are not directly related to the microscopic theory of the spontaneous spin reorientation in rare-earth orthoferrites and orthochromites. For instance, the authors of the most recent paper\,\cite{Luxem} did not take into account a number of interactions, such as the fourth-order anisotropy for the $ 3d $ sublattice of orthoferrites and the crystal field for $ R $-ions, which play a fundamental role in determining the spontaneous spin reorientation. The spin anisotropy of the second order in the $3d$ sublattice of orthorhombic orthoferrites and orthochromites is generally not reduced to an effective uniaxial form as adopted in Ref.\,\cite{Luxem}. Furthermore, the density functional theory does not allow in principle to give an adequate description of such effects of higher orders of perturbation theory as spin anisotropy or antisymmetric exchange\,\cite{CM-2019}.

In this paper, we present the results of a simple but realistic microscopic model of the spontaneous spin reorientation in rare-earth orthoferrites and orthochromites, which takes into account all the main relevant interactions. This model was developed back in the 80s of the last century\,\cite{thesis}, but has not been published until now.

\section{Model formulation}

The most popular examples of systems with the spontaneous SR transitions are magnets based on $3d$ and $4f$ elements such as rare-earth orthoferrites \textit{R}FeO$_3$, orthochromites \textit{R}CrO$_3$, intermetallic compounds \textit{R}Co$_5$, \textit{R}Fe$_2$ etc. In all cases, an important cause of the spontaneous SR is the $4f-3d$ interaction. Usually this interaction is taken into account by introducing an effective field of the magnetically ordered $3d$ sublattice acting on the $4f$ ions.

To consider the contribution of the rare-earth sublattice to the free energy at low temperatures, we are developing a model which takes into account either the well isolated lower Kramers doublet of the $4f$ ions (with an odd number of the $4f$ electrons) or the well isolated two lower Stark sublevels with close energies that form a quasi-doublet.

Within the framework of such ``single-doublet'' approximation we consider the spontaneous SR transition in orthorhombic weak ferromagnets \textit{R}FeO$_3$ and \textit{R}CrO$_3$, where the free energy per ion can be represented as follows
\begin{equation}\label{phiRFEO3}
 \Phi (\theta) = K_1 \cos 2 \theta + K_2 \cos 4 \theta - k T \ln 2 \, \cosh \dfrac{\Delta (\theta)}{2 k T},
\end{equation}
where $K_1$, $K_2$ are the first and second anisotropy constants of the $3d$ sublattice, which are temperature independent (at least in the SR region), $\theta$ is the orientation angle of the antiferromagnetic, or N\'{e}el vector $\bf G$ of the $3d$ sublattice (e.g. in the $ac$ plane), and $\Delta (\theta)$ is the lower doublet (quasi-doublet) splitting of the $4f$ ion in a magnetic field induced by the $3d$ sublattice. 

Theoretical estimations\,\cite{thesis,M_2021,JETP-1981} of the different
contributions to the first constants of the magnetic anisotropy for orthoferrites \textit{R}FeO$_3$ point to a competition of several main mechanisms with relatively regular (Dzyaloshinskii-Moriya (DM) coupling, magnetodipole interaction) or irregular (single-ion anisotropy, SIA) dependence on the type of \textit{R}-ion. For instance, the microscopic theory predicts an unexpectedly strong increase in  values of the constant $K_1(ac)$ for LuFeO$_3$ as compared with YFeO$_3$.
The SIA contribution to $K_1(ac)$ partially compensates for the large contribution of the DM interaction in YFeO$_3$,  whereas in LuFeO$_3$, they add up. 
This result is confirmed by experimental data on the measurement of the threshold field $H_{SR}$ of spin reorientation $\Gamma_4\rightarrow\Gamma_2$ ($G_x\rightarrow G_z$) in the orthoferrite Lu$_{0.5}$Y$_{0.5}$FeO$_3$, in which $H_{SR}$\,=\,15\,T as compared to $H_{SR}$\,=\,7.5\,T in YFeO$_3$\,\cite{JETP-1981}.
Thus, one can estimate $K_1(ac)$ in LuFeO$_3$ as around three times as much as $K_1(ac)$ in YFeO$_3$.

Let us pay attention to recent works on the determination of the parameters of the spin Hamiltonian in YFeO$_3$ from measurements of the spin-wave spectrum by the inelastic neutron scattering\,\cite{Hahn,Park} and terahertz absorption spectroscopy\,\cite{Amelin}. However, these authors started with a simplified spin-Hamiltonian that took into account only Heisenberg exchange, DM interaction, and single-ion anisotropy. Obviously, disregarding the magnetic dipole and exchange-relativistic anisotropy, the ``single-ion anisotropy'' constants found by the authors are some effective quantities that are not directly related to the SIA.

Unfortunately, despite numerous, including fairly recent, studies of the magnetic anisotropy of orthoferrites, we do not have reliable experimental data on the magnitude of the contributions of various anisotropy mechanisms.

As shown by theoretical calculations\,\cite{thesis,M_2021,cubic} the constants $K_2$ of the fourth order spin anisotropy rather smoothly decrease in absolute value, changing by no more than two times on going from La to Lu. But the most interesting was the conclusion about the different signs of these constants, positive for the $ac$ and $bc$ planes and negative for the $ab$ plane, thus indicating a different character of spin-reorientation transitions in the corresponding planes, i.e., second-order transitions in the $ac$ and $bc$ planes and first-order transitions in the $ab$ plane\,\cite{KP}. 
Indeed, all currently known spin-reorientation transitions of the $\Gamma_4 - \Gamma_2$ ($G_x - G_z$) type
in orthoferrites \textit{R}FeO$_3$ (\textit{R} = Pr, Nd, Sm, Tb, Ho, Er, Tm, Yb) 
are smooth, with two characteristic temperatures of the second-order phase transitions to be a start and finish of the spin reorientation, and the only known jump-like first order SR transition for these crystals is the SR transition $\Gamma_4 - \Gamma_1$ ($G_x - G_y$) in the $ab$ plane in DyFeO$_3$\,\cite{KP}. 
A unique example that confirms the conclusions about the sign of the second anisotropy constant is a mixed orthoferrite Ho$_{0.5}$Dy$_{0.5}$FeO$_3$\,\cite{KP}
in which two spin-reorientation
transitions $G_x - G_y$ ($T$\,=\,46\,K) and $G_y - G_z$ (18\,$\div$\,24\,K) are realized through one phase transition of the first order in the $ab$ plane and two phase transitions of the second order in the $bc$ plane, respectively.

The splitting value $\Delta (\theta)$ for the Kramers doublet in a magnetic field $\bf H$ has the well-known form
\begin{equation}\label{DeltaKramers}
 \Delta (\theta) = \mu_{B} \left[ \left( g_{xx} H_x + g_{xy} H_y \right)^2 + \left( g_{xy} H_x + g_{yy} H_y \right)^2 + g_{zz}^2 H_z^2 \right]^{1/2},
\end{equation}
where it is taken into account that for the $4f$ ions in \textit{R}FeO$_3$ the $\hat{g}$-tensor (with the local symmetry $C_s$) has the form 
\begin{equation}\label{g-tensor}
 \hat{g} = 
 \begin{pmatrix}
  g_{xx} & g_{xy} & 0\\
  g_{xy} & g_{yy} & 0\\
  0 & 0 & g_{zz}
\end{pmatrix}.
\end{equation}

The effective field $\bf H$ for the SR transition $G_x\rightarrow G_z$ in the $ac$ plane can be represented as follows
\begin{equation}\label{MagnField}
 H_x = H_x^{(0)} \cos \theta, \,\, H_y = H_y^{(0)} \cos \theta, \,\, H_z = H_z^{(0)} \sin \theta,
\end{equation}
so in the absence of an external magnetic field, for $\Delta (\theta)$ we have the rather simple expression:
\begin{equation}\label{DeltaKramers2}
 \Delta (\theta) = \left( \dfrac{\Delta_a^2 - \Delta_c^2}{2} \cos 2 \theta + \dfrac{\Delta_a^2 + \Delta_c^2}{2} \right)^{1/2},
\end{equation}
where $\Delta_{a,c}$ are the doublet splitting for the cases of $\theta = 0$ ($G_z$-phase) and $\theta = \pi /2$ ($G_x$-phase) respectively. The dependence $\Delta (\theta)$ from \eqref{DeltaKramers2} is also valid in the case of quasi-doublet.

A contribution of splitting $\Delta$ to the free energy $\Phi (\theta)$ for the rare-earth sublattice is usually considered in the ``high-temperature'' approximation, when $kT \gg \Delta$ and the influence of the $4f$ sublattice are reduced only to renormalization of the first anisotropy constant $K_1$:
\begin{equation}\label{Ku*}
 K_1^* = K_1 \left( 1 - \dfrac{1}{\tau} \right),
\end{equation}
where $\tau = T/T_{SR}$ is the reduced temperature and $T_{SR} = (\Delta_a^2 - \Delta_c^2)/ 16k K_1$ is the characteristic transition temperature.

Below we will consider a specific situation when $K_1 > 0$ and $\Delta_a > \Delta_c$, i.e. when the configuration $G_x$ ($\theta = \pi/2$) is realized at high temperatures and a decrease in temperature can lead to the spin reorientation $G_x \rightarrow G_z$ or $G_x \rightarrow G_{xz}$ (transition to an angular spin structure). The type of the phase transition of the spin reorientation in the ``high-temperature'' approximation is determined by the sign of the second constant $K_2$: at $K_2 < 0$ it will be realized by one first-order phase transition at $T = T_{SR}$, i.e. $\tau = 1$, or at $K_2 > 0$ by two second-order phase transitions at $\tau_s = (1 + \gamma)^{-1}$ and $\tau_f = (1 - \gamma)^{-1}$, where $\tau_s$ and $\tau_f$ are the reduced temperatures of the beginning and end of the SR phase transition and $\gamma = 4 K_2 / K_1$.

\section{Analysis of the ``single-doublet'' model}

\begin{figure}[H]
\centering
  \includegraphics[width=0.7\textwidth]{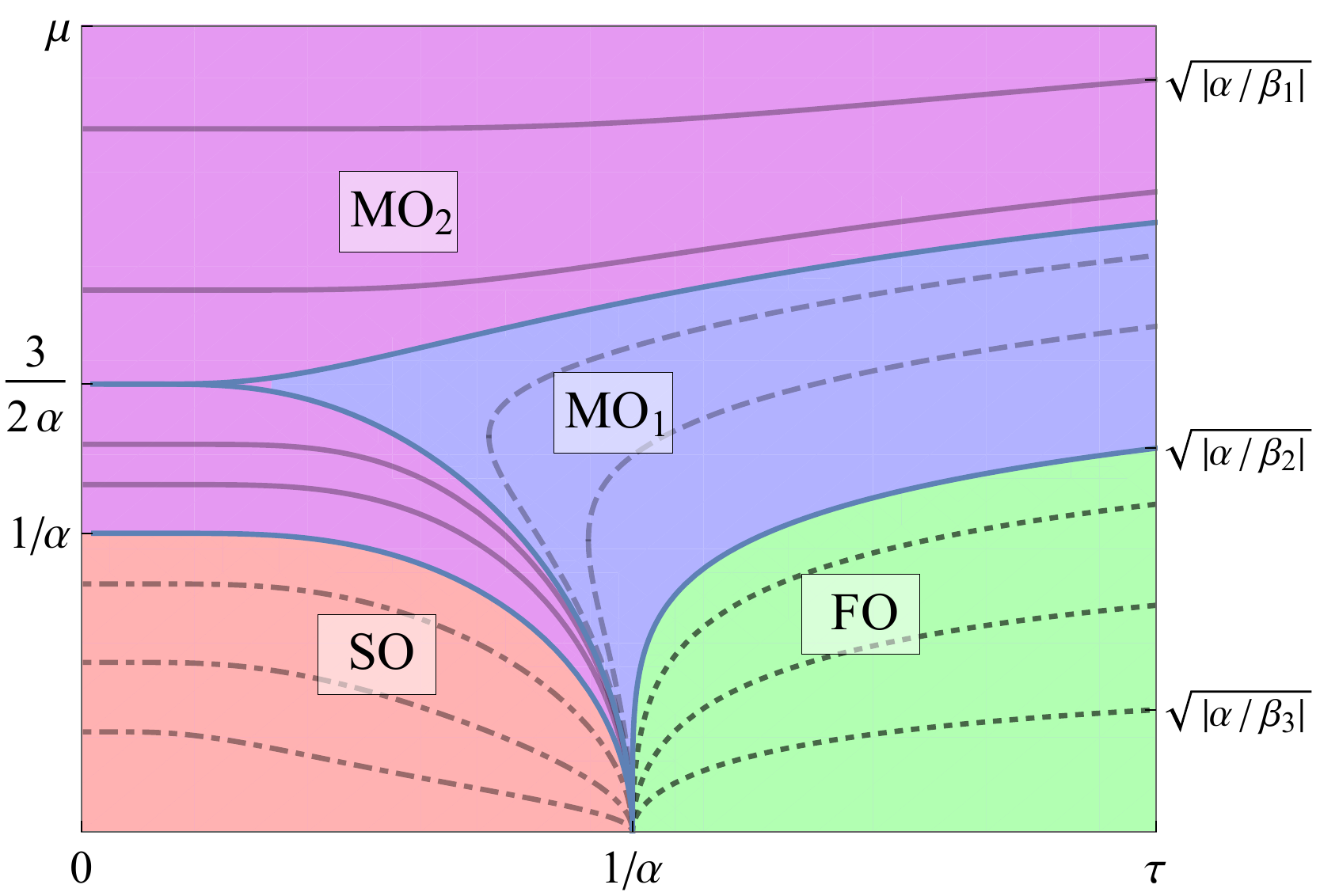}
\caption{(Color online) The peculiar $\mu$-$\tau$ phase diagram which represents solutions of the master equation \eqref{mainEq} given a fixed value of the $\alpha$ parameter and different value of the $\beta$ parameter (see text for detail). }
\label{fig1}
\end{figure}

A behavior of a system described by the free energy \eqref{phiRFEO3} can be analyzed rigorously. The condition $\partial \Phi / \partial \theta = 0$ reduces in this case to two equations:
\begin{equation}\label{SinEq}
 \sin 2 \theta = 0,
\end{equation}
\begin{equation}\label{mainEq}
 \alpha \mu + \beta \mu^3 = \tanh \dfrac{\mu}{\tau};
\end{equation}
where the following notations are introduced:
\begin{equation}\label{mainEq_params}
\alpha = 1 - \gamma \dfrac{\Delta_a^2 + \Delta_c^2}{\Delta_a^2 - \Delta_c^2},\ \beta = \dfrac{2 \gamma}{\mu_f^2 - \mu_{s}^2},\ \mu = \dfrac{\Delta (\theta)}{2k T_{SR}},\ \mu_{s} = \dfrac{\Delta_c}{2k T_{SR}},\ \mu_f = \dfrac{\Delta_a}{2k T_{SR}}.
\end{equation}
This corresponds to three possible magnetic configurations:
\begin{itemize}
\item The configuration $G_x$: $\theta = \pm \pi/2$, stable at $\tanh {\mu_s}/{\tau} \leq \alpha \mu_s + \beta \mu_s^3$ .
\item The configuration $G_z$: $\theta = 0, \ \pi$, stable at $\tanh {\mu_f}/{\tau} \geq \alpha \mu_f + \beta \mu_f^3$ .
\item The angular configuration $G_{xz}$: the temperature dependence of $\theta (\tau)$ is determined by solving the equation \eqref{mainEq} (see Figure\,\ref{fig1}), the state is stable at $\partial \mu / \partial \tau \leq 0$.
\end{itemize}
The peculiar $\mu$-$\tau$ phase diagram which represents solutions of the master equation \eqref{mainEq} given a fixed value of the $\alpha$ parameter and different value of the $\beta$ parameter is shown in Figure\,\ref{fig1}, where areas with different character of the SR transition are highlighted in different colors. For the solutions in the FO region, the SR goes through one first-order phase transition, in the SO region we arrive at one or two second-order phase transitions, in the MO$_{1,2}$ regions we arrive at a “mixture” of the first and second-order phase transitions. All the lines $\mu (\tau)$ on the right side converge to $\sqrt{|\alpha/\beta|}$ at $\tau \rightarrow \infty$; on the left side, when $\tau \rightarrow 0$ the branch point $\mu = \frac{3}{2\alpha}$ is obtained at $\beta = - \frac{4}{27}\alpha^3$,  and the point $\mu = 1/\alpha$ at $\beta = 0$; all the solutions, where $\mu$ can reach zero, converge to $\tau = 1/\alpha$.

The character of the SR transition will be determined by the form of the solution of the equation \eqref{mainEq} in the region $\mu_s \leq \mu \leq \mu_f$. Let us analyze this equation starting with the simplest case $K_2 = 0$, i.e. $\alpha = 1$, $\beta = 0$. In this case, the main equation transforms into the molecular field equation well known in the basic theory of ferromagnetism:
\begin{equation}\label{MolFieldEq}
 \mu = \tanh \dfrac{\mu}{\tau} = B_{\frac{1}{2}} \left( \dfrac{\mu}{\tau} \right) ,
\end{equation}
where $B_{1/2}(x)$ is the Brillouin function. The equation has only one non-trivial solution at $0 \leq \tau \leq 1$, $0 \leq \mu \leq 1$, and the function $\mu (\tau)$ has the usual ``Weiss'' form. Thus, with the absence of the cubic anisotropy ($K_2 = 0$) in the ``single-doublet'' model the SR will be realized either through two second-order phase transitions at $\mu_f \leq 1$ (the complete spin-reorientation $G_x \rightarrow G_z$), or through one second-order phase transition at $\mu_f > 1$, but in this case the SR will be incomplete, i.e. it will end with a transition to the angular spin structure $G_{xz}$. The spin reorientation will begin at a temperature $T_s \leq T_{SR}$ and $T_s$ is equal to $T_{SR}$ only in the case $\mu_s = 0$ ($\Delta_c = 0$), which can be realized in the general case only for Ising ions (e.g. Dy$^{3+}$ in DyFeO$_3$). For this type of ions, the temperature dependence of the ``order parameter'' $\mu$ (in fact the splitting $\Delta (\theta)$ of the doublet) in a close range of $T_{SR}$ will be very sharp: $\mu (T) \sim (T - T_{SR})^{-1/2}$. Nevertheless, the SR will be continuous and the temperature range of the SR $\Delta T = T_s - T_f$ at $\mu \ll 1$ can theoretically reach arbitrarily small values.

Thus, the results of the rigorous analysis of the ``single-doublet'' model are fundamentally different from the conclusions of the simplified model (the ``high-temperature'' approximation), according to which for $K_2 = 0$ the spin reorientation always occurs as the first-order phase transition at $T = T_{SR}$.

For a positive second anisotropy constant ($K_2 > 0$, $\beta > 0$), the main equation \eqref{mainEq} has one non-trivial solution in the region $0 \leq \tau \leq 1/\alpha$, $0 \leq \mu \leq \mu_0$ at $\alpha > 0$, and one in the region $0 \leq \tau \leq \infty$, $ \sqrt{|\alpha/\beta|} \leq \mu \leq \mu_0$ at $\alpha \leq 0$, where $\mu_0$ is determined from the solution of the equation $\alpha \mu_0 + \beta \mu_0^3 = 1$. The situation in this case is very similar to the previous one, i.e. the beginning of the SR will always be a second-order phase transition, and the reorientation will be complete ($G_x \rightarrow G_z$) or incomplete ($G_x \rightarrow G_{xz}$). Note that under the condition $(\mu_f^2 - \mu_s^2)/(\mu_f^2 + \mu_s^2) \geq \gamma$, i.e. $\alpha \leq 0$, the width of the reorientation region becomes very large, even if $\mu_s$ differs slightly from $\mu_f$.
 
For Ising ions at $\Delta_c = 0$, the SR beginning temperature is determined in exactly the same way as in the ``high-temperature'' approximation $T_s = T_{SR} / (1 + \gamma)$.

For a negative second anisotropy constant ($K_2 < 0$, $\beta < 0$), the several fundamentally different solutions of the main equation \eqref{mainEq} are possible. For $K_2^* \geq K_2$, where $K_2^*$ is determined from the condition $\beta = -\frac{1}{3} \alpha^3$, i.e.
\begin{equation}\label{ConditionForK*}
 \dfrac{2 \gamma}{\mu_f^2 - \mu_{s}^2} = -\dfrac{1}{3} \left( 1 - \gamma \dfrac{\mu_f^2 + \mu_{s}^2}{\mu_f^2 - \mu_{s}^2}  \right)^3,
\end{equation}
there is one non-trivial solution of the equation \eqref{mainEq} in the region $1/\alpha \leq \tau < \infty$, $\mu \leq \sqrt{\alpha / \beta}$, but here $\mu (T)$ decreases with decreasing temperature, i.e. $\partial \mu / \partial \tau > 0$. This solution is unstable and there is no fundamental possibility for a smooth rotation of spins, the SR is always realized through the first-order phase transition.

In the intermediate range of values $K_2$ ($K_2^* < K_2 < 0$ or $-\frac{1}{3} \alpha^3 < \beta < 0$) the main equation has two non-trivial solutions, and for one of them $\partial \mu / \partial \tau > 0$ (corresponding to bigger values of $\mu$), and for the second $\partial \mu / \partial \tau < 0$ (corresponding to smaller values of $\mu$). It is convenient to consider separately three areas of variation $\beta$.

1. $-\frac{4}{27} \alpha^3 < \beta < 0$:\\
a) the first solution:  $0 \leq \tau < \infty, \,\, \mu_\text{\scriptsize >} \leq \mu < \sqrt{|\alpha/\beta|}$,\\
b) the second solution:  $0 \leq \tau \leq 1/\alpha, \,\, 0 \leq \mu \leq \mu_\text{\scriptsize <}$,\\
where $\mu_\text{\scriptsize >}$, $\mu_\text{\scriptsize <}$ are the bigger and smaller positive solution of the equation $\alpha \mu + \beta \mu^3 = 1$.

2. $\beta = -\frac{4}{27} \alpha^3$:\\
a) the first solution:  $0 \leq \tau < \infty, \,\, 3/(2 \alpha) \leq \mu < \sqrt{|\alpha/\beta|}$,\\
b) the second solution:  $ 0 \leq \tau \leq 1/\alpha, \,\, 0 \leq \mu \leq 3/(2 \alpha),$\\
moreover, in this case we have a branch point of the main equation solution at $\tau = 0$, $\mu = 1$.

3. $-\frac{1}{3} \alpha^3 < \beta < -\frac{4}{27} \alpha^3$:\\
a) the first solution:  $\tau_0 \leq \tau < \infty, \,\, \mu_0 \leq \mu < \sqrt{|\alpha/\beta|}$,\\
b) the second solution:  $\tau_0 \leq \tau \leq 1/\alpha, \,\, 0 \leq \mu \leq \mu_0$,\\
where the quantities $\mu_0$, $\tau_0$ correspond to the branch points of the main equation solutions.

\begin{figure}[H]
\centering
  \includegraphics[width=1.0\textwidth]{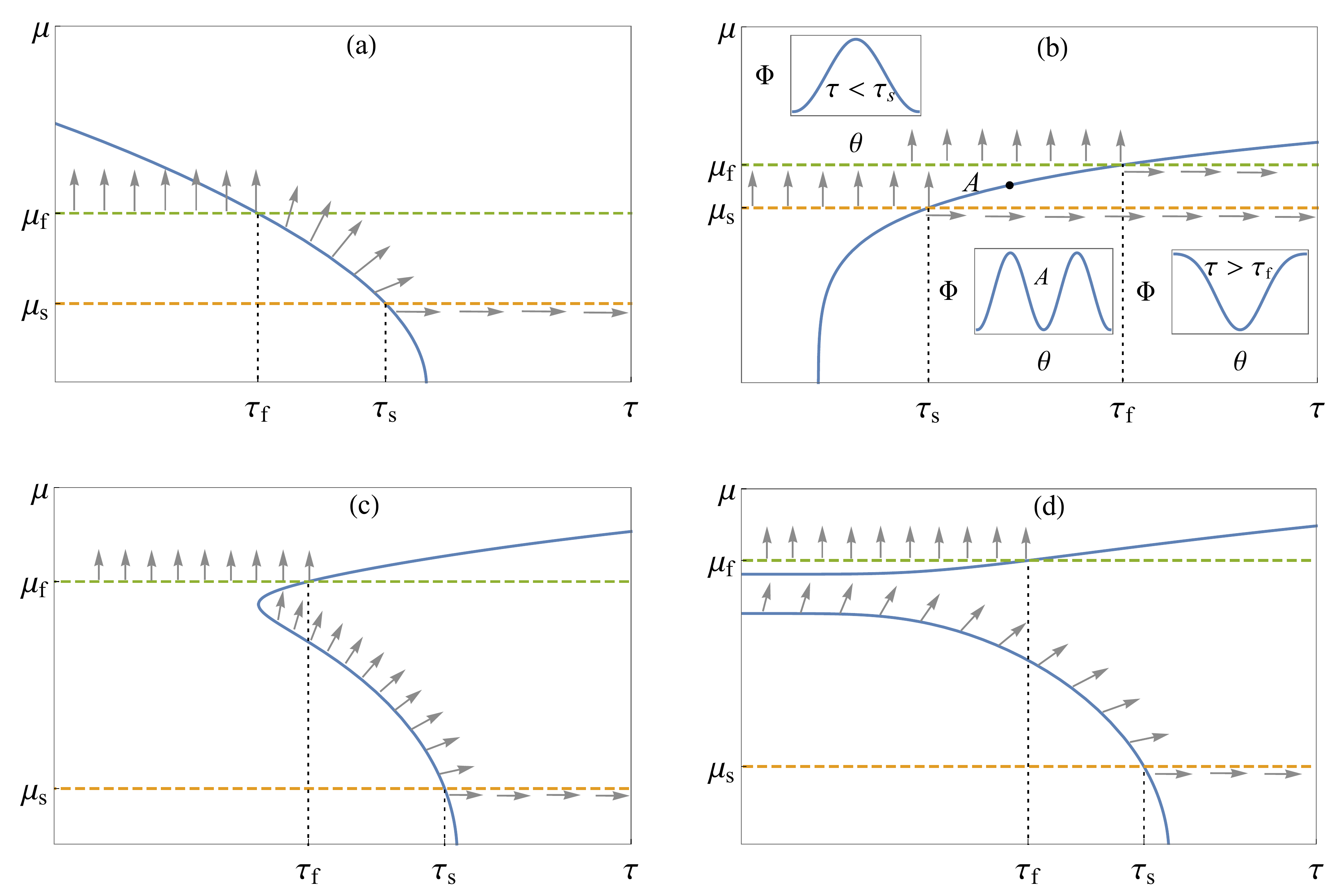}
\caption{Illustrations of typical (a,b) and unconventional (c,d) SR transitions predicted by simple (quasi)doublet model (see text for detail). The arrows indicate the direction of the antiferromagnetic vector $\bf G$ in the $ac$ plane. The insets in panel (b) show the $\theta$-dependence of the free energy.}
\label{fig2}
\end{figure}

Illustrations of typical (a,b) and unconventional (c,d) SR transitions predicted by simple (quasi)doublet model are shown in Figure\,\ref{fig2}. The Figure\,\ref{fig2}a, built with $K_1 = 1,\ \gamma = 0.05,\ \Delta_a = 30.84,\ \Delta_c = 14.82$, which corresponds to $T_{SR} = 45.73,\ \mu_s = 0.162,\ \mu_f = 0.337,\ \tau_s = 1.04,\ \tau_f = 0.91$, describes a typical smooth SR transition
 with two second-order phase transitions $G_x-G_{xz}$  at the beginning ($\tau_s$) and $G_{xz}-G_{z}$ at the end ($\tau_f$) of the spin reorientation.
 
The Figure\,\ref{fig2}b, built with $K_1 = 1,\ \gamma = -0.1,\ \Delta_a = 33.19,\ \Delta_c = 27.1$, which corresponds to $T_{SR} = 22.95,\ \mu_s = 0.59,\ \mu_f = 0.72,\ \tau_s = 0.762,\ \tau_f = 0.93$, describes an abrupt first-order  SR transition. For $\tau > \tau_f$ there is the $G_x$-phase, which can remain stable up to $\tau_s$ when cooled. For $\tau < \tau_s$ there is the $G_z$-phase, which can remain stable up to $\tau_f$ when heated. The point $A$ marks a phase transition point when the phases $G_x$ and $G_z$ have equal energies.

The Figure\,\ref{fig2}c, built with $K_1 = 1,\ \gamma = -0.222,\ \Delta_a = 6.72,\ \Delta_c = 1.63$, which corresponds to $T_{SR} = 2.65,\ \mu_s = 0.307,\ \mu_f = 1.266,\ \tau_s = 0.778,\ \tau_f = 0.523$ and the Figure\,\ref{fig2}d, built with  $K_1 = 1,\ \gamma = -0.25,\ \Delta_a = 6.71,\ \Delta_c = 2.02$, which corresponds to  $T_{SR} = 2.56,\ \mu_s = 0.396,\ \mu_f = 1.31,\ \tau_s = 0.73,\ \tau_f = 0.545$ describe unconventional "mixed" SR transitions. At $\tau_s$ there is the smooth second-order phase transition $G_x-G_{xz}$. At $\tau \leq \tau_f$ we have two stable phases $G_z$ and $G_{xz}$: at those temperatures the sharp first-order phase transition $G_{xz}-G_{z}$ can happen, or the system could stay in the angular $G_{xz}$-phase.

Thus, there are not only the smooth and abrupt SR transitions, a characteristic feature of the range of intermediate values $K_2$ is the fundamental possibility of the existence of ``mixed'' SR transitions, in which the spins first smoothly rotate through a certain angle and then jump to the position with $\theta = 0$. For this, it is sufficient that $\mu_f$ corresponds to a point on the upper branch of solutions, and $\mu_s$ to a point on the lower branch of solutions at $\tau_f < \tau_s$. In this case, the spin reorientation begins with the single second-order transition $G_x \rightarrow G_{xz}$ and then ends with the first-order phase transition $G_{xz} \rightarrow G_z$.
In contrast to the ``high-temperature'' approximation,  the ``single-doublet'' model claims the nature of the phase transition is determined not simply by the sign of the second anisotropy constant, but also it depends on the ratio between $K_1$, $K_2$ and the doublet splitting in both phases. Nevertheless, if we apply the simplified model to describe the SR transition, we have to renormalize both the first and the second anisotropy constant, giving the last one sometimes a rather complicated temperature dependence, in particular with a change in sign when considering transitions of the ``mixed'' type. Of course, in this case Fe sublattice alone is not enough to provide the value of the effective second constant.

\section{Conclusion}
The model of the spin-reorientation transitions induced by the $4f-3d$ interaction in rare-earth orthoferrites and orthochromites has been investigated. It is shown that both the temperature and the character of the spin-reorientation transition following from the solution of the  transcendental equation \eqref{mainEq} are the result of competition between the second and fourth order spin anisotropy of the $3d$ sublattice, the crystal field for 4f ions, and the $4f-3d$ interaction. At variance with the ``high-temperature'' approximation,  the ``single-doublet'' model, along with typical smooth and abrupt SR transitions, predicts the appearance of mixed-type SR transitions, with an initial second-order transition  and a final abrupt first-order transition.

\ \\
\textbf{Funding:} The research was supported by the Ministry of Education and Science of the Russian Federation, project № FEUZ-2020-0054, and by Russian Science Foundation, project № 22-22-00682.

\end{document}